# Effective separation of photogenerated electron-hole pairs by radial field facilitates ultrahigh photoresponse in single semiconductor nanowire photodetectors

Shaili Sett[1,#] and A. K. Raychaudhuri[2,*]

[1]Department of Condensed Matter and Material Sciences, S. N. Bose National Centre for Basic Sciences, JD Block, Sector 3, Salt Lake, Kolkata 7000106, India.

[2]CSIR- Central Glass and Ceramic Research Institute
196 Raja S.C. Mullick Road, Jadavpur, Kolkata-700 032, India.

#Current Address: Department of Physics, Indian Institute of Science, Bangalore-560012, India.
*Email: arupraychaudhuri4217@gmail.com

**Abstract**

We report an investigation on the observation of ultrahigh photoresponse (photogain, $G_{Pc}$ >$10^6$) in single nanowire photodetectors of diameter < 100 nm. The investigation which is a combination of experimental observations and a theoretical analysis of the ultrahigh optical response of semiconductor nanowires, has been carried out with emphasis on Ge nanowires. Semiconductor nanowire photodetectors show a signature of photogating where $G_{Pc}$ rolls-off with increasing illumination intensity. We show that surface band bending due to depleted surface layers in nanowires induces a strong radial field (~ $10^8$ V/m at the nanowire surface) which causes physical separation of photogenerated electron-hole pairs. This was established quantitatively through a self-consistent theoretical model based on coupled Schrodinger and Poisson Equations. It shows that carrier separation slows down the surface recombination velocity to a low value (< 1 cm/s) thus reducing the carrier recombination rate and extending the recombination lifetime by few orders of magnitude. An important outcome of the model is the prediction of $G_{Pc}$ ~ $10^6$ in a single Ge nanowire (with diameter 60 nm), which matches well with our experimental observation. The model also shows an inverse dependence of $G_{Pc}$ on the diameter that has been observed experimentally. Though carried out in context of Ge nanowires, the physical model developed has general applicability in other semiconductor nanowires as well.

# TOC GRAPHICS

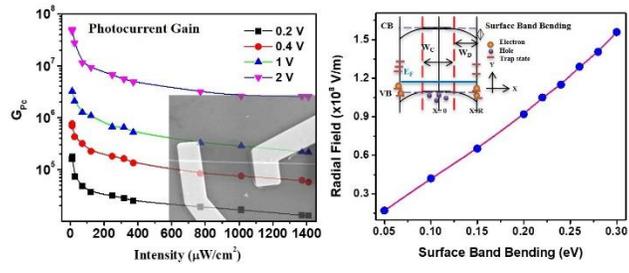

## I. INTRODUCTION

Semiconductor nanowire (NW) photodetectors have been well studied in the past decade due to their high Responsivities which enable detection of low radiation powers.[1-7] The Responsivity ($\mathcal{R} \equiv \frac{I_{Ph}}{P}$ ratio of photocurrent ($I_{Ph}$) to absorbed power ($P$)) in single NW photodetectors reaches a value well in excess of $10^6$ A/W, that is at least six orders of magnitude more than that observed in commercially available p-n junction based detectors (which are workhorses for most applications) made from bulk semiconductors that generally have a peak $\mathcal{R} \leq 1$A/W.[8] These single NW photodetectors are usually photoconductive detectors and have metal-semiconductor-metal (MSM) device configuration with the diameter ($d$) < 100 nm and active length $l$ (distance between electrodes) of few µm. Large values of $\mathcal{R}$ have been observed in a wide class of materials which includes elemental semiconductors like Si[2] and Ge,[7] compound semiconductors like GaAs[6], GaN[5], InAs[4], oxide semiconductors like ZnO,[3, @] and even in molecular semiconductor Cu:TCNQ[9]. In recent years, it has been shown that high $\mathcal{R}$ can also be obtained in an array of nano/microlines of suspended Si made on a standard Silicon-on-Insulator (SOI) wafers with MSM configuration[10,11]. The suspension leads to inhibition of carrier recombination at the Si/SiO$_2$ interface in SOI wafers. Table 1 has a compilation of published results on parameters of photodetectors made from single semiconductor NWs. The origin of the high response may indeed have a common physics that is present in almost all semiconductor NWs, although the quantitative extent may vary depending on the material specificity.

The origin of such high $\mathcal{R}$ has been explored to a certain extent and it has been pointed out that the surface can be the natural choice for the origin of such large photoconductive gain ($G_{Pc} \equiv \frac{N_{carrier}}{N_{photon}}$, the number of carriers per photon absorbed ) that has general applicability in all NWs. An early suggestion (in the context of ZnO NWs[3] which shows $G_{Pc} \sim 10^8$ ) was presence of hole-trap states on the surface, and an *internal electric field* that prevents carrier recombination, thereby increasing photocarrier lifetime ($\tau_{life}$). This was followed by Si NW photodetectors showing phototransistive gain, where trapped carriers at the surface act as an optically modulated gate and increases $\tau_{life}$ through spatial charge separation.[12] This phenomenon where one type of photogenerated carrier gets trapped (due to impurity, interface states etc.) which prolongs their lifetime and creates a spatial distribution was later termed as the "photogating effect".[7,13-17] It may be considered as an applied gate voltage since it modulates the channel conductance which leads to enhancement of $G_{Pc}$.

Foot Notes
@ Note that the ZnO NWs have diameter between 150 – 300 nm.

Another early suggestion[14] was that an electrostatic effect, wherein minority carriers (generated by illumination) are trapped at the surface which induce majority carriers in the NW core/central region. This leads to accumulation of majority carriers in the central channel region, enhancing the conductivity which in turn enhances $G_{Pc}$.

More recently, surface band bending (SBB) in NWs has been suggested as an important enabling factor that enhances photoresponse.[17] The photogenerated minority carriers are localized in the surface depletion region created by SBB, and the excess majority carriers are left behind in the channel leading to enhanced conduction that enhances $G_{Pc}$. The role of SBB and *internal electric field* which leads to charge separation has been qualitatively discussed in the previous reports but a quantitative understanding of the band bending and how the field enhances photoresponse has not been explored till now.

Table 1. Peak responsivity of single nanowire photodetectors: A compilation of published results (arranged in chronological order of publication).

| Nanowire Material | Diameter (nm) | Operating Wavelength (nm) | Working Bias | Peak Responsivity $\mathscr{R}$ (A/W) | $\mathscr{R}$ (A/W)* scaled to 1V | Year | Ref # |
|---|---|---|---|---|---|---|---|
| ZnO | 150-300 | 390 | 5V | $6.3 \times 10^7$ | $1.26 \times 10^7$ | 2007 | 3 |
| Ge | 50 | 532 | 2V | $2 \times 10^2$ | $1 \times 10^2$ | 2007 | 18 |
| GaAs | 90-160 | 522 | 4V | $8.4 \times 10^3$ | $2.1 \times 10^3$ | 2009 | 6 |
| InAs | 30-75 | 300-1100 | 10V | $4.4 \times 10^3$ | $4.4 \times 10^2$ | 2013 | 4 |
| Si | 80 | 300-1100 | 0.1V | $2.6 \times 10^4$ | $2.6 \times 10^5$ | 2014 | 2 |
| Cu:TCNQ | 30 | 405 | 1V | $8 \times 10^4$ | $8 \times 10^4$ | 2014 | 9 |
| GaN | 180 | 380 | 5V | $7.7 \times 10^1$ | $1.4 \times 10^1$ | 2017 | 5 |
| Vertical Ge | 20 | 1550 | 1V | $2.3 \times 10^1$ | $2.3 \times 10^1$ | 2017 | 15 |
| Ge | 30 | 532 | 0.1V | $10^6$ | $10^7$ | 2018 | 16 |
| Ge | 30 | 300-1100 | 2V | $2.4 \times 10^7$ | $1.2 \times 10^7$ | 2018 | 7 and this work |

Foot Notes
*The observed $\mathscr{R}$ increases with applied bias $V$ which is almost linear for low bias. This has been used for scaling $\mathscr{R}$.

It is interesting to note how an in-built axial field in an MSM photodetector can lead to self-powered photodetection.[19] In such a device, a Schottky barrier ($\varphi_B$) forms at the metal-semiconductor (MS) interface. The $\varphi_B$ formation in NWs is dominated by structural inhomogeneity and presence of surface oxide,[19,20] which unintentionally gives rise to unequal barrier heights at the two MS junctions. Due to asymmetry in the barrier heights, there is formation of an axial electric field. This field separates the photogenerated electron-hole pairs and drives them to the respective electrodes even in the absence of an external bias. The *internal field* due to SBB plays a similar role in charge separation. A quantitative evaluation of this *internal field* is unanswered even after a decade of photoresponse studies in semiconductor NWs. We investigate this issue in this report.

We have recently reported broadband photodetection in single Ge NW that can reach $\mathscr{R} = 10^7$ A/W, at a low illumination intensity of ~10 µW/cm$^2$ and at an applied bias of 2V (see Table 1). Such high Responsivity can lead to detection of radiation power as low as 10$^{-15}$ W since the noise limited current detection in these detectors is $\leq 20$ nA. In this work, we report a comprehensive study of photoresponse on single Ge NW based photo detectors ($d <$ 100 nm) made with MSM configuration that investigates its physical origin. The experimental investigation was combined with a theoretical model which allowed us to suggest quantitatively the physical origin. The strong optical response is caused by photogating of the nanowire core (channel region) which in turn is enabled by an effective carrier separation mechanism. The theoretical model based on self-consistent calculation of the coupled Schrodinger-Poisson Equation (SPE) showed existence of a strong radial field with high magnitude (~10$^8$V/m) arising from SBB that enables carrier separation. The analysis showed that the surface recombination velocity becomes exceedingly small which prolongs the carrier recombination time ($\tau_{life}$) by orders of magnitude. Since $G_{Pc} \propto \tau_{life}$, this leads to enhancement of Gain by orders of magnitude. The model also investigates the issue of inverse dependence of $G_{Pc}$ on NW diameter and whether there is any upper limit to $G_{Pc}$ as $d$ is reduced.
Though this work has been carried out in the context of Ge NWs, it is general enough to be applicable to other semiconductor NWs as well.

## II. METHOD

The Ge NWs were grown by Vapor-Liquid-Solid (VLS) mechanism on Si substrates in a multi-zone furnace using physical vapor transport method. The NWs were found to be single crystalline with growth along <111> direction. The NWs are free of oxygen in the bulk albeit with a layer of surface oxide, limiting to ~5 nm thickness. Details of characterization are given in earlier publications from the group.[7,19,21] Transmission electron microscopy (TEM) and High resolution TEM data on Ge NWs are given in Supporting Information (Figure S1).

The single NW photodetectors were fabricated on Si/SiO$_2$ (300nm) wafers. The Ge NWs were connected to Cr/Au contact pads using Cr/Au or Ni metal lines with sub-micron width written using electron beam lithography (EBL) and lift-off or by Focused Electron Beam induced deposition (FEBID) of Pt lines. Image of a typical device is given in Figure 1(a). The details of the physical dimensions and metal contact used in the NW photodetectors are given in Table 2. A total of six single NW devices have been investigated.

The photoconductivity setup had a monochromator coupled to a Xenon lamp source (300-1100 nm). We recorded the photocurrent ($I_{Ph}$) through a source meter with a LabView program to collect the *I-t* and *I-V* curves under illumination and in dark using a GPIB interface. The power falling on the sample was calibrated using a NIST traceable Si detector using a picoammeter. We then use the relation $\mathcal{R} = I_{Ph}/P$, to determine the power (*P*) of the monochromatic light.

### III. RESULTS

In Figure 1(b) we show the response of one of the detectors at a bias of 2V as the illumination is turned ON and OFF at an intensity 550 µW/cm$^2$ (at wavelength $\lambda = 650$ nm). The photocurrent ($I_{Ph}$) is ~ 0.5 µA with a rms noise of $\pm 10$ nA. Even though the dark current in this device (~15 nA) and in some of the devices can be larger, in the range of few tens of nA, there is no component of the leakage current in the device as the leakage in the substrate is < pA. To test the stability of the device, 2000 cycles of illumination ON/OFF was applied, and the photodetector showed no signs of degradation or drift. (Data given in Supporting Information, Fig. S2) The ON/OFF ratio is ~ 10 as shown in Fig 1(b). It is noted that ON/OFF ratio alone does not determine the Responsivity. Other factors such as illumination intensity, rms noise and Specific Detectivity need be accounted for as well. The response time of these MSM devices are generally determined by large stray capacitance arising from the leads which are in the scale of few msec.[22] Figure 1(c) shows the Gain at different applied bias for a NW for intensities in the range of $10^1 - 10^3$ µW/cm$^2$ (details given in Table 2). The Gain is reduced when intensity increases thus showing an inverse relation, which happens due to photogating mechanism and has been explained later. The typical *I-V* curves for devices *F* and *F\** and contour plots of device current as a function of bias with light intensity are provided in the Supporting Information (Fig S3, S4.) The results related to performance parameters on all the single Ge NW photodetectors are summarized in Table 2. It can be seen from Table 2 that $G_{Pc}$ for all the devices (scaled at bias 1V) lie between $10^5$-$10^7$ and in the NW *E* with $d = 30$ nm, $G_{Pc} = 5\times10^7$ is the maximum.

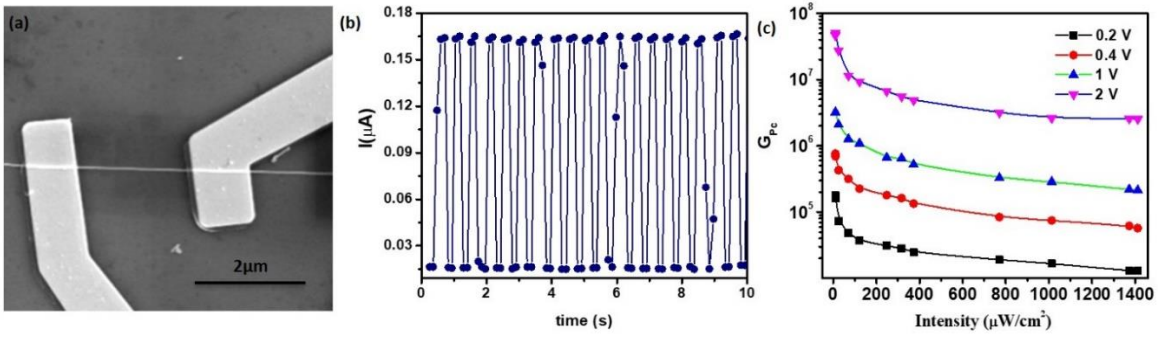

Figure 1. (a) An MSM device of a single NW photodetector connected by metal electrodes. (b) Response of a typical single Ge NW detector for illumination ON and OFF under illumination intensity of 550 µW/cm² taken with $\lambda =$ 650 nm. (b) Gain as a function of Intensity (from 10-1450 µW/cm² ) at different applied voltages. (c) Gain as a function of Intensity (from 10-1450 µW/cm² ) at different applied voltages.

**Table 2.** Performance parameters in single Ge NW photodetectors.

| NW | $d$ (nm) | $l$ (µm) | Bias (V) | Conductivity at RT ($\sigma$) S/cm | Max. $G_{Pc}$ | Scaled[#] $G_{Pc}$ (at 1V) | Intensity (µW/cm²) | $G_N$ (m²/V) |
|---|---|---|---|---|---|---|---|---|
| A* | 90 | 2 | 0 | $10^4$ | 7.1x10⁵ | - | 8.2 | - |
| B | 70 | 1.4 | 0.5 | 5.72 | 1.1x10⁶ | 2.2x10⁶ | 136 | 1.9x10⁻⁶ |
| C | 65 | 3.2 | 1 | 15.38 | 5x10⁶ | 5x10⁶ | 19.4 | 5.1x10⁻⁵ |
| E | 30 | 1.2 | 2 | 20.41 | 5x10⁷ | 2.5x10⁷ | 10.5 | 3.6x10⁻⁵ |
| F | 62 | 1.8 | 3 | 3.33 | 1.0x10⁷ | 3.3x10⁶ | 7.2 | 6.5x10⁻⁶ |
| F* | 62 | 1.8 | 0.1 | $10^2$ | 1.8x10⁶ | 1.8x10⁷ | 960 | 5.8x10⁻⁵ |

Foot Notes
*# The scaled $G_{Pc}$ is measured $G_{Pc}$ divided by bias in V*
*\* The same NW after vacuum annealing*

The internal factors as well as device parameters are linked to the opto-electronic parameters through $G_{PC}$. In a two terminal MSM photoconductive detector the device parameter $G_{PC}$ is given by[23]

$$G_{Pc} = \frac{\tau_{life}}{\tau_{transit}} = \frac{eV}{k_BT}\left(\frac{l_{avg}}{l}\right)^2 \quad (1)$$

where, $k_B$ is the Boltzmann constant, $T$ is the temperature, $\tau_{life}$ is the carrier recombination time, $\tau_{transit} = \frac{l^2}{\mu V}$, is the transit time for carriers diffusing between the electrodes of distance $l$ with mobility $\mu$ under an applied bias of $V$. In case the carrier recombines within an average recombination length $l_{avg}$, the recombination time is given as $\tau_{life} = \frac{l_{avg}^2}{\mu\left(\frac{k_BT}{e}\right)}$. Since the photodetectors have different values of $l$ and measurements are made with different applied bias, to have a scale for comparison, we define a Normalized Gain $(G_N)$, that is dependent only on internal parameters to the NW (such as the mobility) and is given by,[24]

$$G_N \equiv \left(\frac{l^2}{V}\right) G_{Pc}. \quad (2)$$

As shown in Table 2, $G_N$ varies by an order of magnitude from ~5x10$^{-5}$ to ~ 6x10$^{-6}$ $m^2/V$ and it has an increasing trend with the conductivity ($\sigma$). As $\sigma = ne\mu$ where $n$ is the carrier concentration, the enhancement of $G_N$ on enhancement of $\sigma$ can arise from increase of $n$ or $\mu$. To establish that the expected $G_{Pc}$ is enhanced due to enhancement of $\mu$, we have vacuum annealed NW $F$ at 400$^0$C for 10 mins. The annealed sample is termed as $F^*$. Annealing leads to increase of conductivity and reduction in contact resistance both by a factor of nearly thirty. $G_N$ changes by a factor of nine and this additional enhancement arises from reduction in contact resistance. (The contact resistance leads to reduction in effective bias as part of the applied bias $V$ drops at the contact. Thus, the lower contact resistance leads to enhancement of effective bias to its full value.) The photoresponse depends on the conductivity (and hence mobility) and the diameter of the nanowire as well as on surface states. If all other internal parameters (like conductivity etc.) are kept constant, then the photoresponse is enhanced when the diameter is smaller. The variation of $G_N$ with NW diameter is given in Supporting Information (Fig. S5). Similar observations have also been made before.[14]

One of the important characteristics of $G_{Pc}$ is its dependence on illumination intensity ($\mathcal{J}$). We find that this is a general relation where $G_{Pc}$ rolls off at higher values of $\mathcal{J}$ following the relation[3,7]

$$G_{pc}(\mathcal{J}) = \frac{\tau_{life}/\tau_{transit}}{1+\left(\frac{\mathcal{J}}{\mathcal{J}_0}\right)^\xi} \quad (3)$$

where the ratio $(\tau_{life}/\tau_{transit})$ is the photogain for $(\mathcal{J} \to 0)$, $\xi$ is a phenomenological fitting parameter that depends on the recombination of carriers at the trap states. In Figure 2(a) we show $G_{Pc}$ as a function of $\mathcal{J}$ for three representative Ge NW photodetectors along with the fits and Table 3 gives the fit parameters.

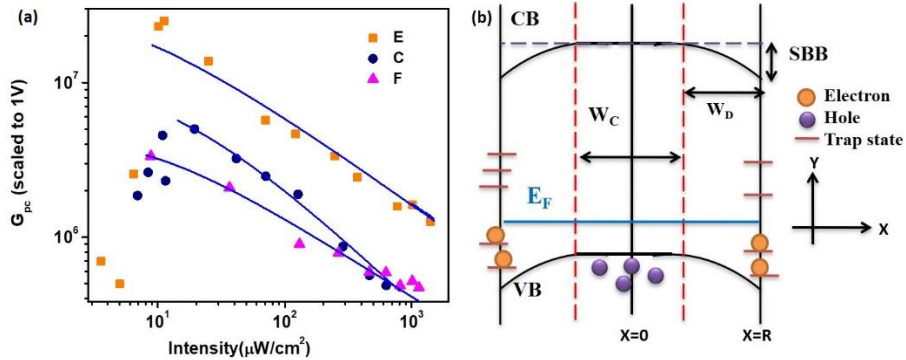

Figure 2. (a) Dependence of the photogain $G_{pc}$ as a function of illumination intensity in NWs E, C and F. The lines through the data are fits to Eqn. 3. (b) Schematic of a p-type nanowire of radius $R\left(=\frac{d}{2}\right)$, showing surface band bending (SBB) with depletion of electrons at the nanowire core and depletion of holes at the surface. The depletion width is denoted as $W_D$ and the Conduction channel width at the NW core as $W_C$. The y-axis is along the axial direction and the x-axis is along the radial direction of the nanowire. The conduction and valence band edges are marked as CB and VB respectively and the Fermi level is indicated by $E_F$.

Such a dependence of $G_{Pc}$ on intensity was first proposed in the context of ZnO NWs[3] and has often been taken as a manifestation of photogating effect[13]. If one type of photogenerated carrier gets trapped by interface states, it inhibits carrier recombination and prolongs the $\tau_{life}$. A schematic of the concept of photogating is shown in Figure 2(b). This aspect will be discussed quantitatively in next section.

Compilation of results on semiconductor NWs (Table 1) and the results on single Ge NW photodetectors presented above (summarized in Table 2) and the clearly show the extent of high $G_{Pc}$ achievable in NW photodetectors. It shows that the enhancement of mobility brought about by vacuum annealing the NW can also lead to a large enhancement in $G_{Pc}$. The results also show the generality of the roll-off of Gain at high illumination intensity following the phenomenological relation given by eqn. 3.

Table 3. Fit parameters of illumination dependence of $G_{Pc}$ to Eqn. 3

| NW Name | $\tau_{life}/\tau_{transit}$ | $\mathcal{J}_0(\mu W/cm^2)$ | $\xi$ |
|---|---|---|---|
| C | $1.0 \times 10^7$ | $9.0 \pm 2$ | 0.73 |
| E | $2.5 \times 10^7$ | $8.3 \pm 1.4$ | 0.60 |
| F | 4.8x10⁶ | 8.8±1.1 | 0.52 |

## IV. DISCUSSION

### A. Gain and length scales

There are two important length scales determined by the physical structure of a photoconductive MSM device that would determine the performance of the device. As shown in Eqns. 1 and 2 these length scales are $l_{avg}$ and $l$. At room temperature and at a moderate bias $V = 1\ V$, $\frac{eV}{k_B T}$ contributes a factor of ~ 40. Thus, if the carriers are effectively collected before they recombine (when $l_{avg} > l$), $G_{Pc}$ can be a large quantity. More conventional bulk photodetectors with $l \sim$ few tens of $\mu m$, ($l_{avg} \ll l$) limits $G_{Pc}$ to a small value. In a very clean NW, that does not have recombination within the bulk of the NW, it is expected that the photogenerated carriers will effectively recombine at the surface of the NW (unless there is slowing down of recombination velocity). In that case, it is expected that $l_{avg} \sim d$ and for a given $l$, as per Eqn. 1 $G_{Pc} \propto d^2$ i.e. it will decrease as the $d$ decreases which is contrary to our observation as well as that made by other investigators.

A simple (and in a manner of speaking straight jacketed) application of Eqn. 1 for NWs without qualification of the recombination mechanism and recombination velocity is that the magnitude of $l_{avg}$ leads to further contradictions as shown below. Using $\frac{k_B T}{e} \approx 25$ meV at room temperature, and using applied bias as typically $1V$, from eqn. 1 we can obtain $l_{avg}$. For $G_{Pc} \approx 10^6$-$10^7$ (and $l \sim 1\mu m$), the value of $l_{avg}$ thus obtained is few hundreds of micrometers. This is non-physical as the value exceeds the physical length of the NW. However, a large $G_{Pc}$ can arise if the $\tau_{life}$ is enhanced without allowing the carriers to diffuse a large distance. This may be possible if there is a *internal field* that physically separates the carriers, as we shall discuss below.

### B. Necessity of a Radial field for separation of photogenerated electron-hole pair

In a conventional photoconductive detector, there is no field that separates out the electron-hole pairs unlike that in junction-based photodetectors where a built-in field at the p-n junction that separates out the carriers. In fact, absence of carrier separation in most conventional photoconductive detectors' limits $G_{Pc}$ to a low value. The built-in field in a junction type photodetector is ~ 1.1x10⁷ V/m for a typical doping of $10^{23}$/m³ and it increases with doping concentration.[25] This field is strong enough to separate out the photogenerated electron-hole pair. Another example in which multiple electron-hole pairs are produced, is through avalanche mechanism due to a breakdown field, where an extremely high electric field exists in the depletion region.[26] In photodetectors made from single NWs of small diameter there arises an *internal field* as has been referred to before[3] which we term the "radial field". Band bending at the surface leads to existence of the radial field that enables separation of photogenerated carriers.

### C. Formation of the radial field

The radial field is related to the SBB due to the depletion layer formed by transfer of charge from interior to surface states or any other states[27] that lie within the band gap of a material. This depletes the surface and pins the Fermi level to the surface, resulting in surface band bending ($\varphi_r$).[28] In a thin NW due to its small diameter ($d < 100$ nm), the depletion width $W_D$ can be a substantial part of the NW diameter and this limits the conducting region in the center of the NW along the axial direction. A schematic is given in Fig. 2(b). There have been observations of SBB in Ge[29] probed by X-Ray Photoelectron Spectroscopy (XPS). The SBB in Ge NWs is approximately 0.3 eV which has been obtained from XPS[22] through the shift in Ge 3d core level.

### D. Framework for using a self-consistent approach to solve Schrodinger-Poisson Equation

In this sub-section we use a self-consistent approach to quantitatively evaluate the important fall outs of SBB. We iteratively solve two coupled equations namely the Poisson equation and the Schrodinger equation till self-consistency is achieved using finite element method.[30,31] The schematic in Figure 3 shows the steps of the simulation.

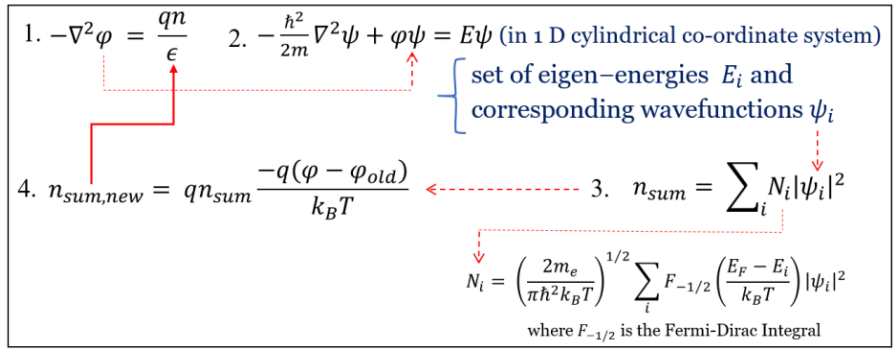

Figure 3. Schematic of the SPE self-consistent solver.

The Poisson's equation, $-\nabla^2 \varphi = \frac{qn}{\epsilon}$ (where $qn$ is the charge density and $\epsilon$ is the dielectric constant of the semiconductor) is first solved to obtain the potential ($\varphi$) created by a given charge distribution ($qn$) in a NW. As a starting point for the charge density, we assume a uniform distribution of dopants in the NW.

The electric potential $\varphi$, obtained from here is then applied as the potential energy term in the Schrödinger equation, $-\frac{\hbar^2}{2m}\nabla^2\psi + \varphi\psi = E\psi$ from which the wave functions ($\psi_i$) and eigenvalues ($E_i$) are calculated. The density function ($n$), which contributes to the space charge density in the electrostatics is determined from these set of $\psi_i$ and $E_i$. The particle density profile $n_{sum}$ is computed using a statistically weighted sum of the probability densities as shown in step 3 of the schematic. Using $n_{sum}$, we can calculate the new charge density $qn_{sum}$. We use an updated charge density $(qn_{sum,new})$[32] to re-estimate an improved $\varphi$ using Poisson equation. The SPEs are solved iteratively until a self-consistent solution is obtained with the condition that the potentials $\varphi$ and the $\varphi_{old}$ differ by less than 1mV (units are in eV) which is the tolerance limit. More details of the method are given in the Supporting Information (Section 3).

### E. Results of Simulation

Figure 4 shows simulation results for SBB potential, radial field ($E_r$), and carrier density in a Ge NW of $d = 60$ nm as a function of distance from the NW central axis (radial distance). The electric potential developed in the NW shows band bending downwards since it is a p-type NW. The magnitude of SBB (as a function of radial distance) has been obtained from the simulation and is shown in Figure 4(a). The corresponding Radial field (see Fig. 4(b)) is zero at the NW center and rises sharply beyond 25 nm from the NW center and has a magnitude equal to $1.55 \times 10^8$ V/m at the surface which is much larger than the typical built-in field of a junction diode.[25] $E_r$ lies within $1.55 \pm 0.02 \times 10^8$ V/m for $d < 150$ nm, beyond which it reduces sharply (see Supporting Information, Fig. S6). The quantitative estimate of such a field has not been done as per our knowledge and its evaluation is important to validate the hypothesis that a sufficiently strong field exists in NWs that can separate out the carriers.

Figure 4(c) shows the carrier density in a 60 nm NW for both types of carriers. Due to presence of acceptor type surface states in p-type Ge,[33-36] the surface (up to ~ 5 nm) is electron rich while the concentration of holes ($n_h$) at the surface is at least five orders less (also see Fig. 6). There is also a concomitant depletion of electrons in the NW core while majority carrier holes are confined towards the bulk of the NW. As compared to electron density ($n_e$) at the NW center, $n_h$ is greater by nearly two orders of magnitude, forming a conducting channel region $W_c$ (as shown as a schematic in Fig. 2(b)). The ratio $\frac{n_e}{n_h}$ is ~ $10^5$ at the NW surface clearly indicating electrons being spatially separated from the holes (as shown in Figure 4 (c)) which gives rise to a large lifetime $\tau_{life}$ thus enhancing $G_{PC}$ (see equation 1).

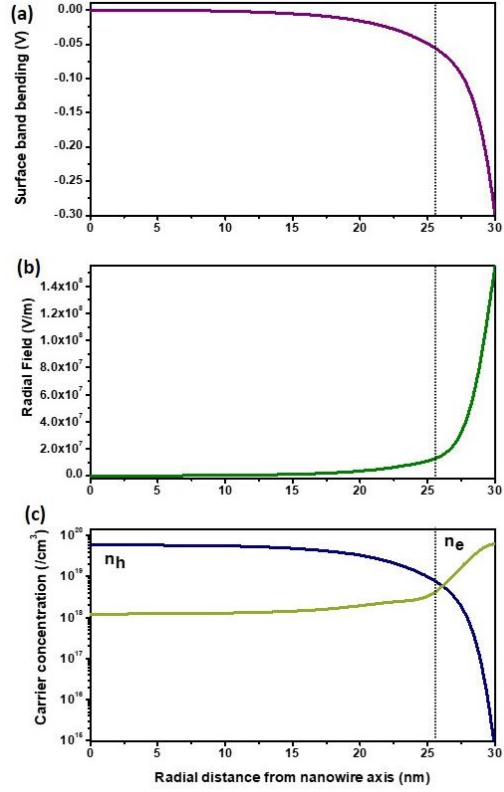

Figure 4. Results of self-consistent solution of SPE in a 60 nm diameter Ge NW. (a) Surface band bending as a function of radial distance along the NW. The SBB ensures a potential = -0.3eV at the surface. (b) The Radial field as a function of radial distance in a 60 nm Ge NW. (c) Carrier distribution along the radial direction showing separation of electrons and holes with confinement of electrons ($n_e$) at the surface (green) while the excess of holes ($n_h$) in the NW channel (blue).

Under illumination, electron-hole pairs are formed and separate out due to SBB. The electrons get trapped at the surface, while holes move to the channel region. With increasing illumination, more traps get filled and the energy bands flatten or SBB is reduced. This leads to free electrons which can now enter the channel region and increases recombination. This also leads to saturation of the photocurrent versus intensity behaviour as the illumination intensity increases thus effectively lowering $G_{PC}$. Our simulation shows that as SBB decreases, so does the radial field which is the cause of separation of the electron-hole pair (see Fig. 5(a)). Lowering of the SBB therefore indicates a condition of the NW under illumination, which causes photogeneration and their spatial separation by the radial field.

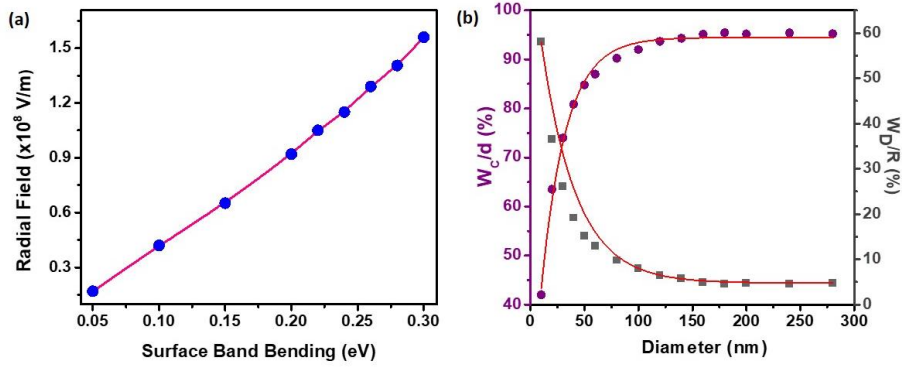

Figure 5. (a) The magnitude of the radial field at the surface as a function of the SBB potential. (b) Variation of ratio Channel width ($W_c$)/ diameter ($d$) (circle) and ratio Depletion width($W_D$) /radius ($R$) (square) as a function of NW diameter for a fixed band bending = 0.3eV.

We define the depletion width ($W_D$) as the distance up to which the radial field lowers to 1/10$^{th}$ of its value at the surface. It is demarcated by a dotted line in Figure 2(b) and 4(c). $W_D$ obtained from the simulation was found to vary from 3.0 nm to 6.7 nm for $10 < d < 280$ nm. The estimated value of the $W_D$ when calculated from the classical relation, $W_D = \sqrt{2\epsilon V/eN_d}$,[37] = 2.9 nm, which is fixed by the doping concentration $N_d$, dielectric constant $\epsilon$ and in-built barrier $V$ (which is the SBB). Nevertheless, it is of similar magnitude as that obtained from simulation. In case of a NW, the distribution of $n_h$ and $n_e$ varies with diameter, causing a variation in $W_D$. The relation given above does not incorporate the spatial variation in $n_h$ and $n_e$. In Fig. 5(b), we show the variation of ratio of the channel width ($W_c = d - 2W_D$) with diameter of the NWs. The $W_c/d$ saturates to 95% beyond 160 nm while for a NW of diameter 30 nm, it reduces to 75%. The same graph (Fig. 5(b)) also shows the variation of ratio of depletion width ($W_D$) to the radius ($R = \frac{d}{2}$). It is observed that in a 200 nm diameter NW, $W_D$ is ~ 5 nm and the conduction channel exist almost throughout the NW (95%). This condition almost mimics a bulk semiconductor system. While for a NW with diameter ~10 nm, it is nearly 60% depleted. This graph quantitatively indicates the impact of SBB on the conduction NWs of low diameters. The conduction channel being restricted, has large impact on the conduction and hence photoconduction of a NW.

### F. Impact on Surface Recombination Velocity

So far, we have established that the formation of SBB and a depletion of holes at the NW surface leads to a radial field of magnitude ~10$^8$ V/m. The inflation of $G_{Pc}$ to ultrahigh values would need enhancement of $\tau_{life}$. Here we show that the SBB reduces the recombination velocity ($v_{reco}$) significantly which leads to extremely long $\tau_{life}$. A study on passivated Si NWs shows surface recombination velocity ($S_{RV}$) ~ 20 cm/s only[38] which is orders of magnitude smaller than that seen in bulk. This aspect of lowering of $S_{RV}$ which can enhance $G_{Pc}$, has not been appreciated before. This is an outcome of the theoretical analysis carried out in this report. The surface recombination velocity is given by,[39]

$$S_{RV} = \Sigma v_{th} N_t, \tag{6}$$

where $v_{th}$ is the thermal velocity ($5\times10^6$ cm/s for Ge), $\Sigma$ is the capture cross-section which is typically $10^{-15}$cm$^2$ [40] and $N_t$ is the number of surface recombination centers. In case of Ge NWs, we may consider the number of trap states $\sim 10^{12}$/cm$^2$ [41,42] which gives $S_{RV} = 5\times10^3$cm/s. The carrier lifetime ($\frac{1}{\tau_{life}} = \frac{4 S_{RV}}{d}$)[43] is calculated 50 ns at the surface. This value is too small to justify any enhancement of $G_{Pc}$. However, it has been shown that the electron-hole recombination rates not only depend upon the surface trap states $N_t$ as shown in Eqn. 6, but on surface band bending as well.[40,44] The recombination rate is known to be lower if there is band bending. A physical reason for a low $S_{RV}$ is the presence of a potential/barrier that prevents one type of carrier to reach the surface. This reduces the recombination rate by an amount $e^{-\varphi_r/kT}$.[40]

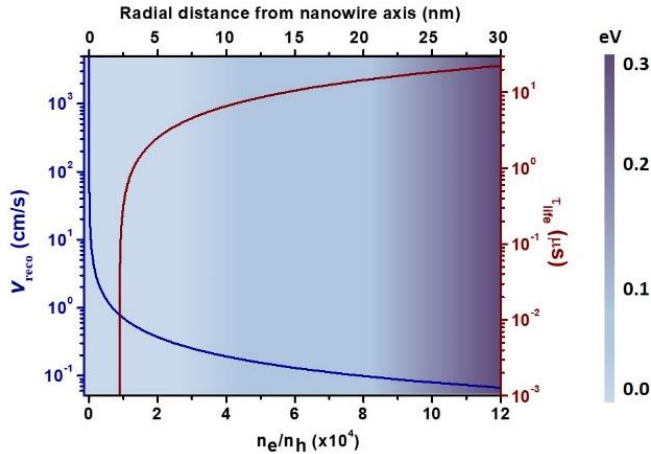

Figure 6. The effect of electron/hole concentration ratio ($\frac{n_e}{n_h}$) on recombination velocity ($v_{reco}$) and carrier lifetime ($\tau_{life}$) along with its variation from the axis of the NW in the radial direction (given by the top x-axis) for a Ge NW of $d = 60$ nm. The color code on the right shows Surface Band Bending.

In Figure 6, the double x-axis shows the electron to hole carrier density ratio ($\frac{n_e}{n_h}$) as well as radial distance from the NW axis. The double y-axis has $v_{reco}$ and $\tau_{life}$, which varies with the $\frac{n_e}{n_h}$ ratio and radial distance, while the color plot shows the SBB. For the case of Ge NWs which has strong band bending (~ 0.3 eV), the recombination from bulk reduces by a factor of $10^5$ at the surface, while in the core of the NW, where SBB is zero, there is no effect on the recombination rate. This reduces the recombination rate by the same factor and enhances $\tau_{life}$ and hence $G_{Pc}$ by five orders of magnitude. As the ratio is enhanced on separation and localization of electrons near the surface, the recombination rate reduces significantly and at the surface of the NW, $v_{reco} = 0.07$ cm/s which gives $\tau_{life} = 22$ μs. Low values of $v_{reco}$ ranging from 0.005 – 0.2 cm/s has been observed in solar cells which have a built-in field.[45] This particular result is one of the most important finding of the theoretical analysis presented in this work.

We have seen from previous publications[7] as well as this work that $\tau_{life}/\tau_{tran} \sim 10^7$. For a device with $l$ ~1 μm, and bias $V = 1V$, the approximate $\tau_{transit} = \frac{l^2}{\mu V} = 1.67 \times 10^{-11}$ sec with $\mu = 600$ cm$^2$/Vs.[33] With lifetime of $\tau_{life} \approx 22$ μs estimated from Figure 6, we get $G_{Pc} = 1.25 \times 10^6$ which matches well with our experimental observations. This quantitatively establishes the efficacy of the approach and establishes that low $v_{reco}$ can be an enabling factor for enhanced $G_{Pc}$ which in turn arises from the separation of the carrier by the radial field created by SBB. This also establishes the main hypothesis proposed in the paper.

### G. Dependence of Gain on diameter

The important experimental observation on NW photodetectors is that $G_{Pc}$ has an inverse dependence on the diameter $d$ provided factors such as mobility are unchanged. A suggestion[14] based on electrostatics of trapped minority carrier at the surface was proposed to explain this inverse dependence on $d$. Here we show that the SBB and the resulting effective carrier separation that controls $v_{reco}$ can lead to an effective inverse dependence of $G_{Pc}$ on $d$.

Even for a constant value of SBB, it is observed that the ratio of $n_e$ at the surface ($x = R$) to $n_h$ at the centre of the NW ($x = 0$) enhances with decreasing $d$ and the effect is more pronounced, smaller the diameter. A larger value of $n_e$ (trapped electrons at the surface) in comparison to $n_h$ in the channel region would lead a larger $W_D$ and higher $G_{Pc}$ because of prolonged carrier life time caused by effectively trapped minority carriers in the depletion layer. Thus, a smaller diameter leads to larger $G_{Pc}$. However, there is a limit to the enhancement of $G_{Pc}$ as diameter is reduced. For NWs with $d < 10$ nm, it is almost completely depleted since $W_D \to \frac{d}{2}$ and $W_c \to 0$ (see Figure 5(b)). In total depletion condition, the carrier concentration $n_h \to 0$ and $\sigma \to 0$. Thus, there would exist a critical diameter $d_c$, below which $G_{Pc}$ would start to decrease with decreasing $d$ after reaching a peak value. It will be an experimental challenge to investigate this keeping mobility and other factors unchanged.

### H. Other factors influencing the Gain

In addition to the main effects arising from SBB and inhibiting recombination by effective carrier separation as discussed before, there are other factors that can also contribute to enhanced photoresponse. One such factor can be impact ionization[46,47]. A radial field strength of ~$10^8$V/m is comparable to the breakdown field in a bulk Ge[24] which is ~4.5x10$^7$ V/m for doping of 10$^{24}$/m$^3$ (as used in our study; see Supporting Information, Section 3.). This process would lead to creation of multiple electron-hole pairs by avalanche mechanism.[26,46] Even though the depletion region is small (~ few nm), it can multiply the electron-hole pair as they gain enough energy to commence the avalanche process.

Another factor that plays an important role is the drift velocity of holes which allows the photodetector to perform in the drift limited regime. In such high fields, the drift velocity may saturate. For example, the saturation drift velocity of a nominally doped Ge wafer is ~6x10$^8$ m/s at 300K at $E = 10^6$V/m.[48] Thus, there is acceleration of photogenerated carriers to their saturation drift velocity at such high fields.

These factors mentioned above in synergy with the main mechanism would enable ultrahigh photoresponse in single semiconductor NWs.

## V. CONCLUSIONS

In summary, we established that ultrahigh photogain $G_{Pc}$ and photoresponsivity $\mathcal{R}$ is a general characteristic of single NW photodetector with $d < 100$ nm made from different semiconducting materials. Such ultrahigh optical response had been proposed to arise from photogating effect by photogenerated minority carriers that are trapped in the depletion layer at the surface of the NWs. The experimental investigations were carried out on Ge NW (30 nm $\leq d \leq$ 90 nm) with different types of metals contacting the semiconductors to ascertain that the high $G_{Pc}$ as well as $\mathcal{R}$ does not have much dependence on contact electrode material and primarily is a property of the Ge NW. The experiment also enables us to generalize the roll-off of $G_{Pc}$ at higher illumination intensity which is an important characteristic that points to saturation of photogating at higher illumination. We could establish through annealing (that improves the mobility by an order) a direct quantitative connection of $G_{Pc}$ to mobility that is an essential characteristic of MSM type photoconductive detector.

We presented results of calculation using a self-consistent model based on coupled Schrodinger-Poisson Equation, which was carried out to on a p-type Ge NW with a surface band bending of 0.3 eV. It quantitatively estimates the physical effects of the surface band bending and showed that it leads to a radial field of magnitude ~10$^8$ V/m (depending on the extent of band bending) which is large enough to move photogenerated electron-hole pairs into separate physical spaces. It leads to the minority carrier (electrons) being separated in the depletion width near the surface causing accumulation of holes in the conduction channel. At the surface the ratio of electron and hole density can be as large as 1.4x10$^5$. The important outcome of the theoretical investigation is the observation that such effective carrier separation slows down the surface recombination velocity to a low value < 1 cm/sec. This in turn reduces the carrier recombination rate thereby extending the recombination lifetime $\tau_{life}$ to tens of microseconds. This enhances the $G_{Pc}$ which is $\propto \tau_{life}$ and the model estimates that in a NW with diameter of 30 nm, $G_{Pc} > 10^6$ can be reached. The model also proposes an inverse dependence of $G_{Pc}$ on the diameter above a critical value.

**Acknowledgement**

AKR acknowledges financial support from Science and Engineering Research Board (SERB), Government of India as a SERB Distinguished Fellow (project number: SB/DF/008/2019).

**Supporting Information**

Supporting Information includes TEM study of Ge NW, Experimental plots of photodetectors, simulation details for the SPE and related graphs. This material is available free of charge via the Internet at http://pubs.acs.org.

# Supporting Information

**Effective separation of photogenerated electron-hole pairs by radial field facilitates ultrahigh photoresponse in single semiconductor nanowire photodetectors**

Shaili Sett[1,#] and A. K. Raychaudhuri[2] *

[1]*Department of Condensed Matter and Material Sciences, S. N. Bose National Centre for Basic Sciences, JD Block, Sector 3, Salt Lake, Kolkata 7000106, India.*

[2] *CSIR- Central Glass and Ceramic Research Institute
196 Raja S.C. Mullick Road, Jadavpur, Kolkata-700 032, India.*

#Current Address: Department of Physics, Indian Institute of Science, Bangalore-560012, India.
*Email: arupraychaudhuri4217@gmail.com

## 1. Transmission Electron Microscopy (TEM) images of Ge nanowire

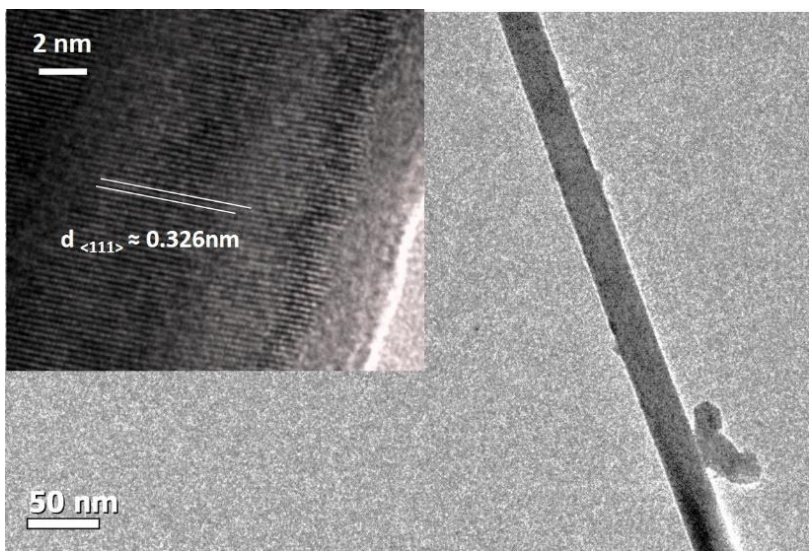

Figure S1. TEM image of a single Ge NW suspended in a Cu grid. A high resolution TEM image showing parallel d planes along <111> direction. The natural surface oxide layer ~ 2 -3 nm.

## 2. Typical photoresponse experimental plots

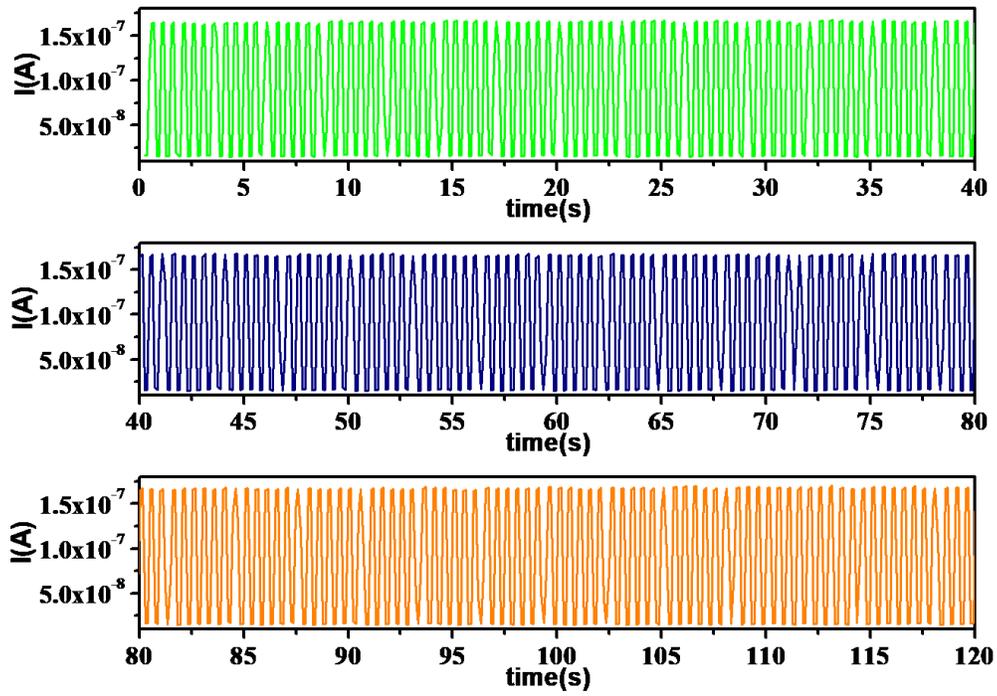

Fig S2. Current versus time in a Ge NW photodetector with illumination ON and OFF for 2000 cycles showing stability of device.

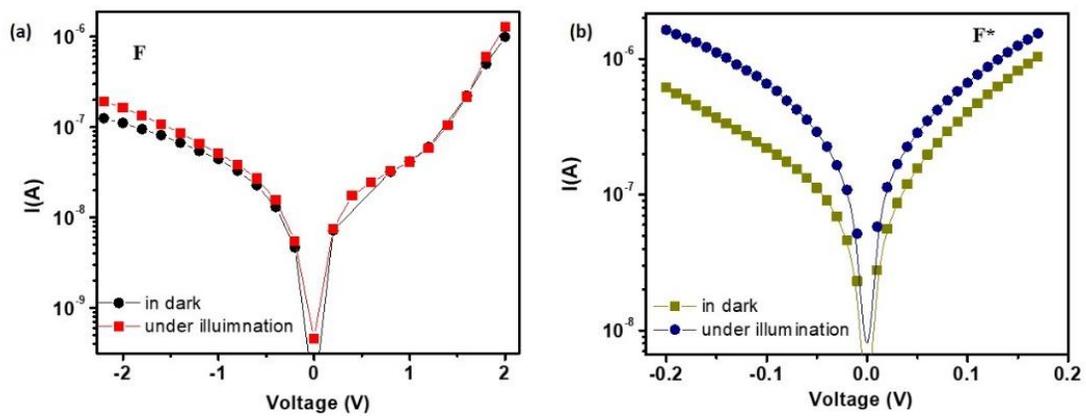

Fig. S3. I-V curves in dark and under illumination for NW (a) F and (b) F* (nanowire F renamed after annealing).

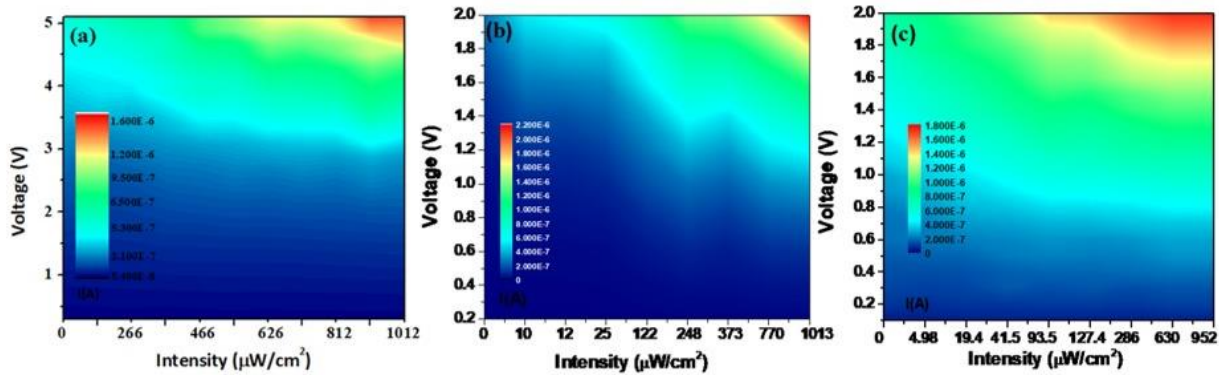

Fig. S4. I-V curves in dark and under illumination for NW F, E and C. Reprinted in part with permission from Sett, S.; Ghatak, A.; Sharma, D.; Kumar, G. V. P.; Raychaudhuri, A. K. Broad Band Single Germanium Nanowire Photodetectors with Surface Oxide-Controlled High Optical Gain. *J. Phys. Chem. C,* 2018, **122**, 8564. Copyright 2018 American Chemical Society.

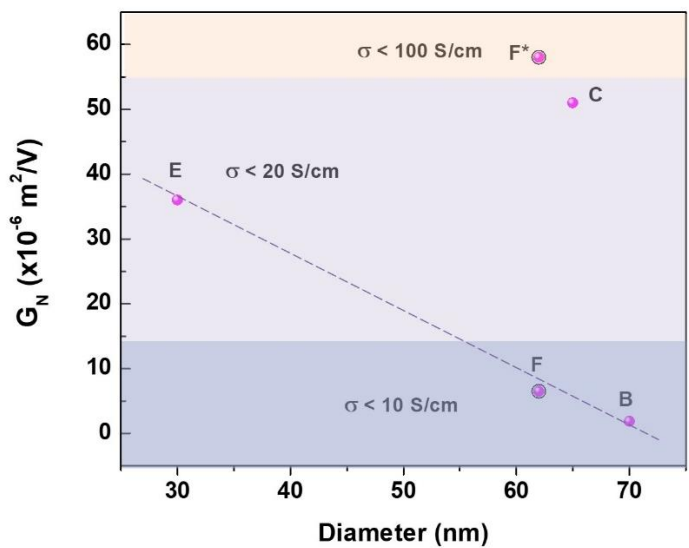

Fig S5. Normalized Gain as a function of NW diameter. The inverse dependence of response on the diameter is apparent. Two nanowires NW C and annealed NW F* show higher response because they have higher conductivity.

## 3. Finite Element Method to solve the self-consistent Schrodinger-Poisson Equation (SPE)

We solve for the system of Germanium nanowire (NW) of diameter 60 nm in 1D geometry using cylindrical symmetry. We assume an infinitely long cylindrical NW, azimuthally symmetric with x-axis taken as the radial direction. At the center of the NW, x = 0. We have used the following parameter sets for the simulations:

- There is uniform background of dopants which are ionized at room temperature. We have taken $N_d$ = $10^{18}$/cm$^3$ which we determine from the average resistivity of ~$10^{-2}$Ωcm.[1]

- We assume the Fermi level ($E_F$) is at 0 eV at the center of a NW (where x = 0).

- Potential at the surface ($\varphi_r$) is pinned below 0.3 eV with respect to $E_F$ since it is known from XPS measurements that the SBB is ~ 0.3 eV in p-type Ge NWs as discussed in section D.3 of manuscript.

- We find the solution at a temperature of 300 K.

- Bulk Ge material properties like dielectric constant etc. has been used.

- The SPE solves for only one type of carrier. The spatial distribution of the other carrier, in this case electron, can be found using the relation, $n = N_c e^{\frac{E_F - E_c}{k_B T}}$,[2] where $N_c$ is the density of states in the conduction band and $E_F - E_c$ corresponds to the potential $\varphi_r$, which we have determined from SPE.[32]

After we determine the eigenfunctions $\psi_i$ from the Schrodinger equation, the particle density profile $n_{sum}$ is computed using a statistically weighted sum of the probability densities where,

$n_{sum} = \sum_i N_i |\psi_i|^2$ .

Considering a 1D cylindrical geometry, $N_i$ is given by,[3]

$N_i = \left(\frac{2m_e}{\pi \hbar^2 k_B T}\right)^{1/2} \sum_i F_{-1/2}\left(\frac{E_F - E_i}{k_B T}\right) |\psi_i|^2$

where, $F_{-1/2}$ is the Fermi-Dirac, $k_B$ is the Boltzmann constant, $T$ is temperature, $E_F$ is Fermi level, $m_e$ is effective mass of the carrier and $\hbar$ is the reduced Plank's constant.

## 4. Radial field as a function of Diameter

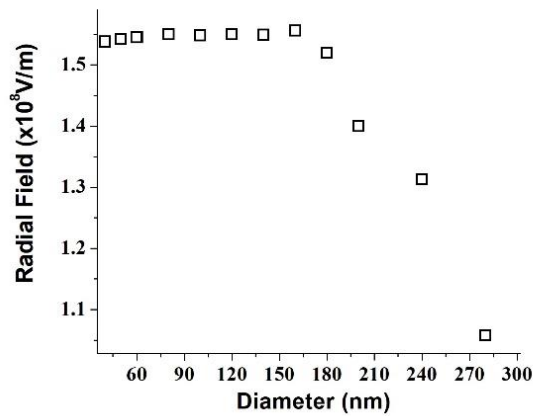

Figure S6. Radial field dependence on the NW diameter.

## References

[1] D. B. Curtiss, *Bell Syst. Techn. J.* **40** 509, 1961

[2] C. Kittel and H. Kroemer, Thermal Physics, (W. H. Freeman, United States of America, 1980)

[3] J. H. Luscombe and A. M. Bouchard, *Phys. Rev. B.* **46**, 263-268, 1992